\documentclass[11pt,twoside]{article}

\usepackage{asp2004}
\usepackage{epsf}
\usepackage{psfig}
\usepackage{graphicx}
\usepackage{lscape}

\begin{document}
\title{Spitzer Studies of Ultracool Subdwarfs: Metal-poor Late-type M, L and T Dwarfs}   
\author{Adam J.\ Burgasser}   
\affil{American Museum of Natural History, Central Park West at 79th Street, New York, NY 10024}    

\begin{abstract} 
Recent discoveries from red optical proper motion and wide-field
near-infrared surveys have uncovered a new population of ultracool
subdwarfs, metal-poor stars and brown dwarfs extending into the
late-type M, L and possibly T spectral classes.  These objects are
among the first low-mass stars and brown dwarfs formed in the
Galaxy, and are valuable tracers of metallicity effects in
low-temperature atmospheres. Like solar metallicity late-type dwarfs,
ultracool subdwarfs emit the majority of their emergent flux at
infrared wavelengths.  Here I discuss how Spitzer observations
will contribute to the study of these objects, enabling measurement
of the temperature scale, tests for dust formation and the identification of
surface gravity and metallicity diagnostics in the coldest brown dwarfs.
\end{abstract}

\section{Ultracool Subdwarfs}

Subdwarfs are metal-deficient stars, classically defined as lying
below the stellar main sequence in optical color-magnitude
diagrams \citep{kui39}. These objects are not subluminous
but rather hotter (i.e., bluer optical colors) than equivalent
mass main sequence dwarfs, a consequence of their reduced metal
opacity (e.g., Chamberlin \& Aller 1951). Cool subdwarfs (spectral types sdK
and sdM) are typically found to have halo
kinematics, and are presumably relics of the
early Galaxy.  They are therefore
important tracers of Galactic chemical history and are
representatives of the first generations of star formation.

The high space velocities of halo subdwarfs allow them to stand out in proper
motion surveys \citep{rei84}.  Early surveys using photographic
plates (e.g., LHS catalog, Luyten 1979a;
APM Proper Motion Survey, Scholz et al.\ 2000) identified several cool subdwarfs
with spectral types as late as sdM7/esdM7 \citep{giz97} and
effective temperatures (T$_{eff}$) down to 3000 K \citep{leg00}.
More recent surveys, such as the SUPERBLINK catalog
\citep{lep03b,lep03d} and the SuperCosmos Sky Survey
\citep[SSS]{ham01a}, have pushed our compendium of cool subdwarfs to
even cooler temperatures.  With spectral types
extending beyond the end of the M dwarf regime \citep{sch04},
these objects are {\em ultracool subdwarfs}.

\section{L and T Subdwarfs}

The first L subdwarf, 2MASS 0532+82 \citep{me03c}, was
serendipitously identified in the 2MASS catalog as a faint source
invisible in optical survey plates (e.g., POSS-II). Its optical
and $J$-band spectrum is similar to that of an L7 dwarf (Figure
1), but its near-infrared colors are uniquely blue due to enhanced
collision-induced H$_2$ absorption (e.g., Saumon et al.\ 1994).
Optical and near-infrared bands of CaH, CrH and FeH are enhanced
in this source, while the 2.3 $\micron$ CO bands are absent. These
are the same molecular abundance patterns seen in M subdwarfs, due
to the preferential formation of single metal hydrides over double
metal oxides in cool metal-poor atmospheres \citep{mou76}. The
high proper motion of 2MASS 0532+82 ($\mu = 2{\farcs}6$ yr$^{-1}$,
implying $V_{tan} \approx 250$ km/s) confirms it as a very cool
(T$_{eff} \leq 2000$ K) halo subdwarf. Four other L subdwarfs and
L subdwarf candidates have been identified in the SUPERBLINK
\citep{lep03a}, 2MASS \citep{me04b}, SSSPM \citep{sch05} and SDSS
\citep{siv05} surveys.

\begin{figure}[th]
\centering
    \includegraphics[width=8.3cm]{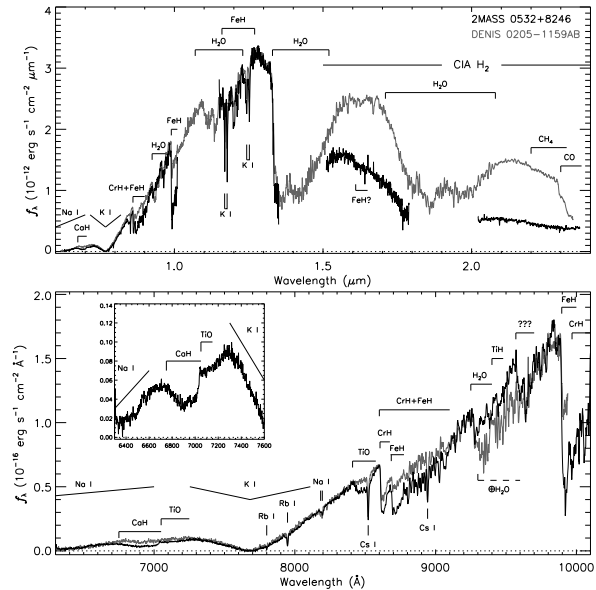}
\caption{{\em Top:}
The 0.6-2.5 $\micron$ spectrum of the L subdwarf 2MASS 0532+82 (black) as compared
to the L7 DENIS 0205-11 (grey).  While
red optical and $J$-band features are similar, the near-infrared
spectrum of 2MASS 0532+82 is distinctly blue due to enhanced H$_2$
absorption.  {\em Bottom:} Red optical spectra of 2MASS 0532+82 and DENIS 0205-11.
Note the enhanced metal hydride bands in
2MASS 0532+82, consistent with subsolar metallicity; and
the unexpected strength of TiO bands, perhaps indicative of
suppressed condensate dust formation (from Burgasser et al.\ 2003b).
~\label{fig1}}
\end{figure}

2MASS 0532+8246 is so cool that it is very likely a brown dwarf.
Because halo stars are old in general, the vast majority of halo
brown dwarfs should also be old and have cooled to T dwarf
temperatures (T$_{eff}$ $\leq$ 1500 K).  Indeed, a T subdwarf
population may be dominant in the halo \citep{me04a}.  Currently,
only one source is a viable T subdwarf candidate, 2MASS 0937+29
\citep{me02a}.  This T$_{eff}$ $\approx$ 900 K brown dwarf has
blue near-infrared colors ($J-K_s$ = $-$0.6) consistent with
enhanced H$_2$ absorption, and an unusually strong 0.99 $\micron$
FeH band \citep{me03d}.  Comparison of empirical data to spectral
models also supports a subsolar metallicity \citep{bur02}.
However, it remains unclear as to whether surface gravity effects
are responsible for these features \citep{me02a,kna04}, and 2MASS
0937+29 does not exhibit halo-like kinematics \citep{vrb04}.
Spitzer observations of this and other peculiar T dwarfs may help
resolve this issue.

\section{The Potential of Spitzer Observations}

The abundance of molecular gaseous and condensate
species in the atmospheres
of late-type M, L and T dwarfs implies that their infrared spectra
are highly sensitive to temperature, gravity and metallicity
variations.  Comprehensive investigations of these effects
are currently underway (e.g., Gorlova et al.\ 2003).
Spitzer observations of ultracool subdwarfs will be useful
in identifying specific metallicity diagnostics in the
4-20 $\micron$ regime, where several major
gaseous (e.g., H$_2$O, CH$_4$ and NH$_3$) and condensate (e.g., silicates)
species are found.


\noindent {\bf Temperature Scales:}
At wavelengths greater than $\sim$11 $\micron$, the SEDs of ultracool dwarfs
approach the Rayleigh-Jeans approximation for a thermal blackbody
(M.\ Marley, 2005, priv.\ comm.).
Hence, measurement of the absolute flux at these wavelengths yields a direct
determination of temperature ($F_{\lambda} \propto {\rm T}/{\lambda}^4$).
While the temperature scale of
solar-metallicity late-type dwarfs has been studied in detail (e.g., Golimowski
et al.\ 2004), ultracool subdwarf temperatures are largely unknown.
Combining infrared fluxes with forthcoming parallax determinations (F.\ Vrba, 2005, priv.\
comm.) will enable these measurements.


\noindent {\bf Condensate Dust Formation:}
Condensate dust is prevalent in the photospheres of late-type M and L dwarfs
\citep{all01}, and is believed to be
responsible for the red near-infrared colors ($J-K \sim$ 1.5--2.5),
depletion of TiO and VO gases \citep{lod02}, and
photometric variability (e.g., Gelino et al.\ 2002)
observed in these objects.  Yet major condensates such as perovskite
(CaTiO$_3$) and enstatite (MgSiO$_3$) are multiple metal species, and their
formation may be inhibited in metal-poor atmospheres.
There is indirect evidence of this
with the retention of TiO gas bands in the L subdwarf 2MASS 0532+82 (Figure 1).
The absence of dust can be readily traced thought its influence in the
5-9 $\micron$ region.  Here, dust
in M and L dwarf atmospheres produces a smooth, relatively featureless spectrum; without
dust, major absorption bands of H$_2$O and CH$_4$ are prominent (Figure 2).
IRS observations of ultracool subdwarfs can test this prediction and explore the
consequences for atmospheric chemistry.

\begin{figure}[th]
\centering
    \includegraphics[height=4.7cm]{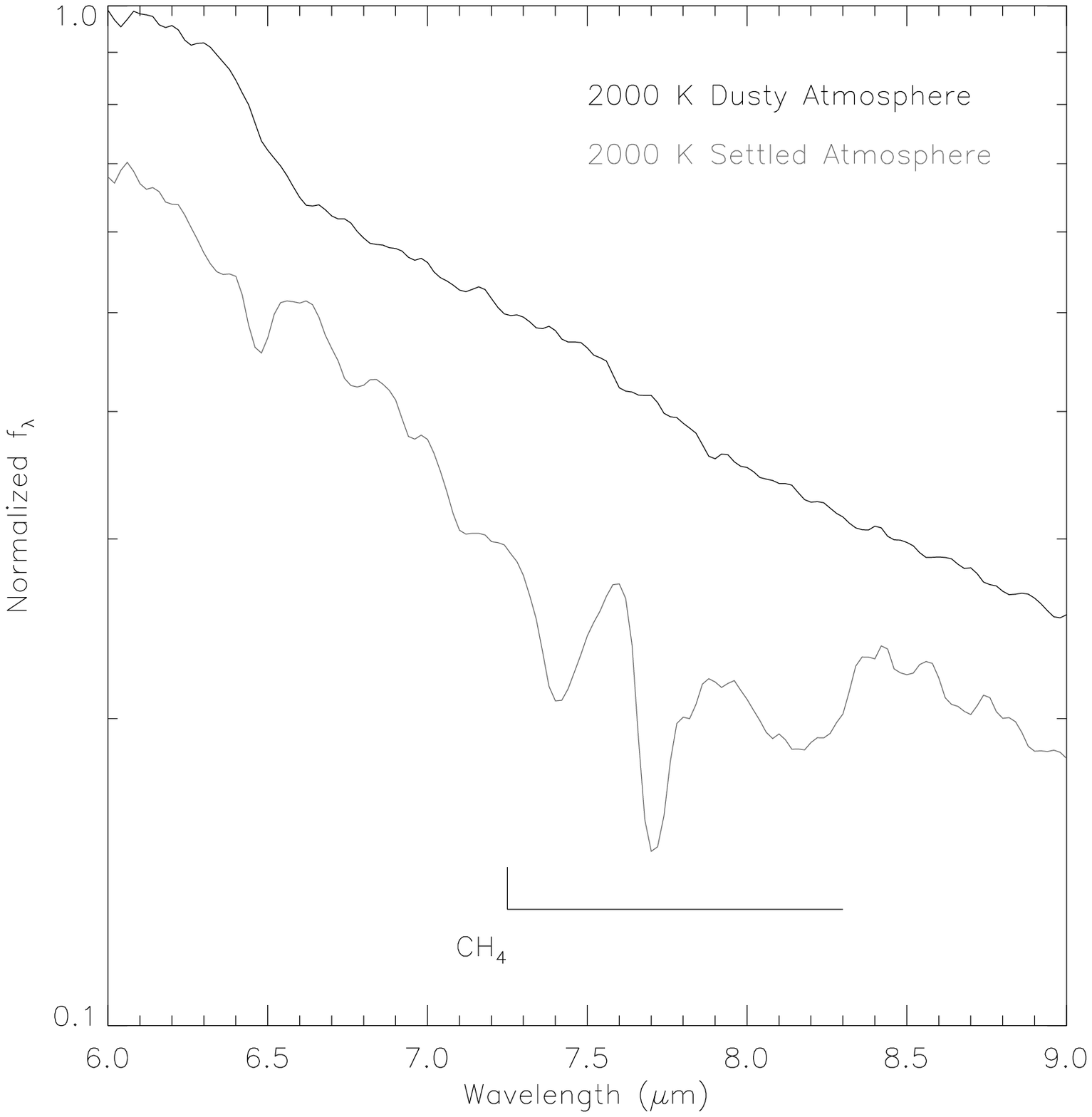}
    \includegraphics[height=4.7cm]{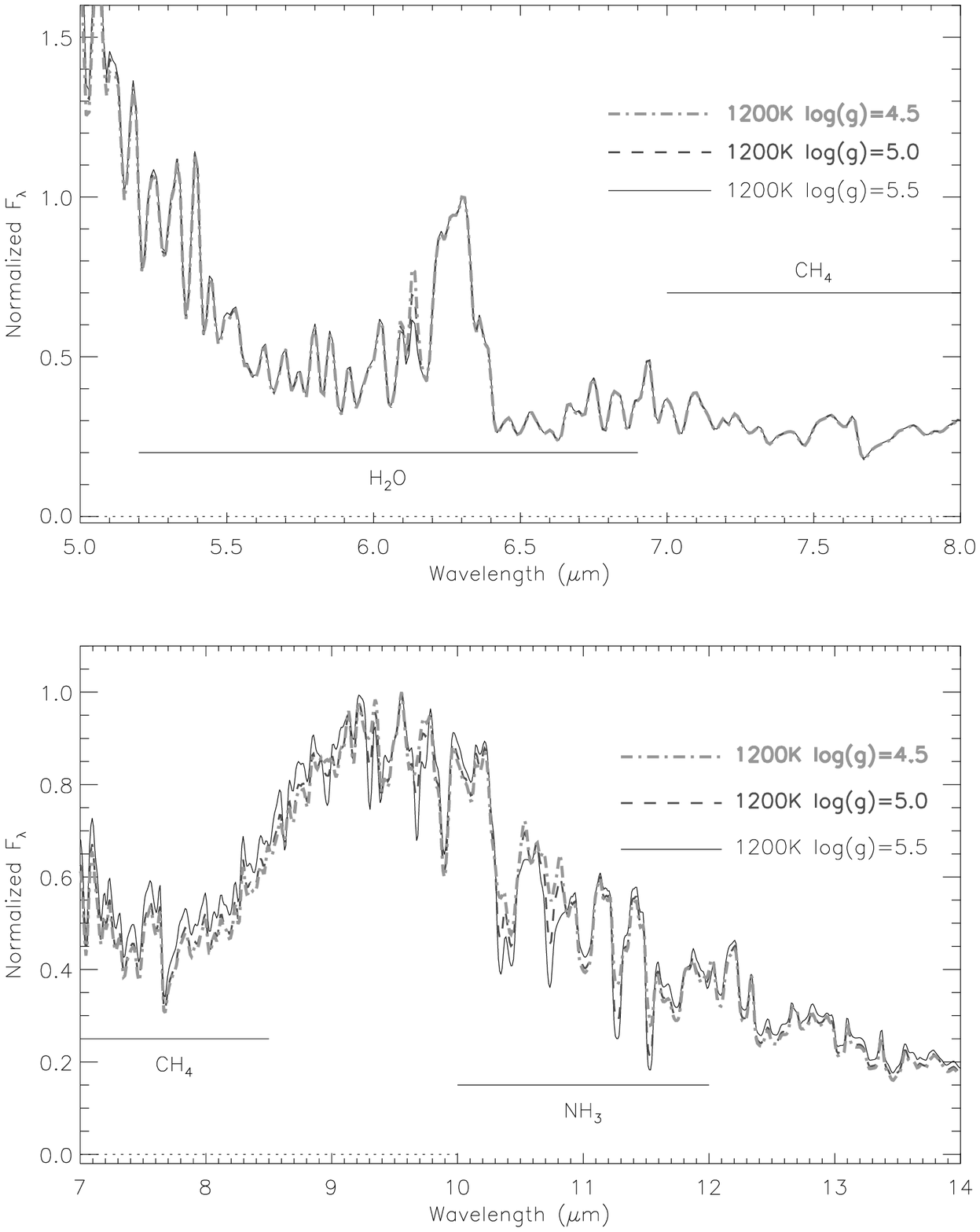}
    \includegraphics[height=4.7cm]{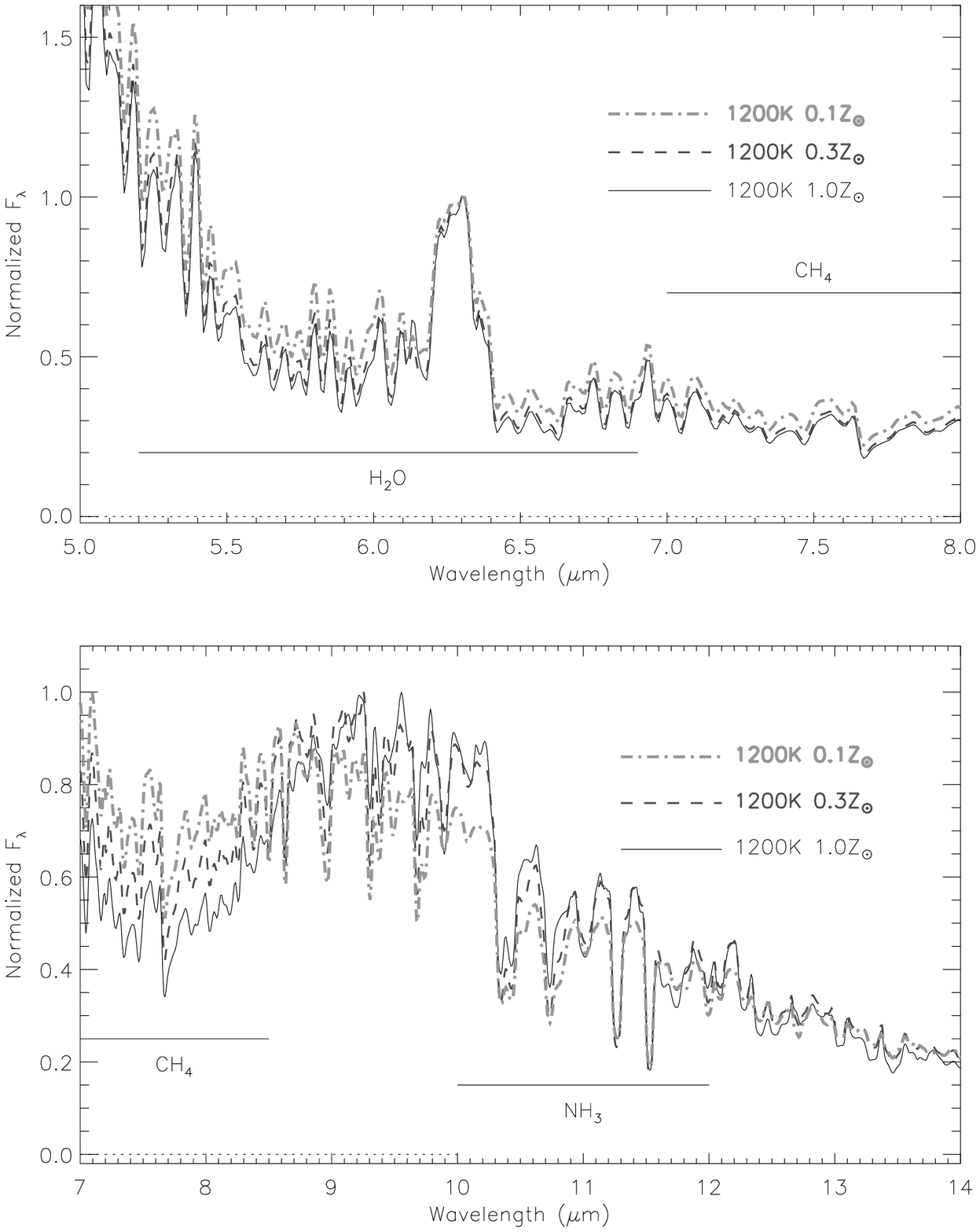}
\caption{{\em Left:} Comparison of 2000 K theoretical spectra in
the 6-9 $\micron$ region with (top) and without (bottom) dust. The
presence of strong molecular features can test whether dust
formation is inhibited in metal-poor atmospheres. {\em Middle and
Right:} Theoretical model spectra for 1200 K brown dwarfs in the
IRS SL2 (top) and SL1 (bottom) spectral windows, normalized at 6.3
and 8 $\micron$, respectively.  The middle panels show variations
due to surface gravity for (top to bottom) $\log{g}$ = 4.5 to 5.5
(cgs) and $Z = Z_{\odot}$. The right panels show variations due to
metallicity for $\log{Z/Z_{\odot}}$ = -2 to 0  and $\log{g} = 5.5$
(models courtesy A.\ Burrows and P.\ Hauschildt).
 ~\label{fig2}}
\end{figure}


\noindent {\bf Gravity versus Metallicity Diagnostics:}
Gravity and metallicity effects in the optical and near-infrared spectra of
T dwarfs produce similar
enhancements in pressure-sensitive K I and H$_2$ features \citep{me02a,me03d,kna04}.
Disentangling these effects is nontrivial.  In the longer wavelength bands
sampled by the IRS spectrograph, gravity effects are minimal, but reduced
metallicities flatten the IR spectra due to increased H$_2$ opacity
and weak metal features (Figure 2).  Empirical confirmation
and characterization of these differences will enable the segregation of old
disk and halo brown dwarfs for future studies of the substellar population
in the vicinity of the Sun.

\acknowledgements The author thanks A.\ Burrows and P.\ Hauschildt
for the use of their theoretical spectral models.  Support for
this work was provided by NASA through the Spitzer Fellowship
Program.


\end{document}